\PassOptionsToPackage{dvipdfmx}{graphicx}  
\documentclass[sigconf]{acmart} 
\usepackage{enumitem}
\AtBeginDocument{%
  }

\renewcommand\footnotetextcopyrightpermission[1]{} 
\begin{document}

\title{The Stage Comes to You: A Real-Time Tele-Immersive System with 3D Point Clouds and Vibrotactile Feedback}

\author{%
Takahiro Matsumoto,
Takahiro Kusabuka,
Hiroshi Chigira,
Kazuhiko Murasaki,
Kakagu Komazaki,
Masafumi Suzuki,
and Masakatsu Aoki
}
\affiliation{%
  \institution{NTT, Inc.}
  \city{Tokyo}
  \country{Japan}
}
\renewcommand{\shortauthors}{Matsumoto et al.}

\begin{abstract}
We present a low-latency tele-immersive entertainment system that streams 3D point clouds and performers’ footstep vibrations, creating the sense that the stage is present.
Moving performers and their surroundings are captured as dynamic point clouds under rapidly changing lighting, then processed, transmitted, and rendered within a total latency of less than 100 ms.
Under high ambient noise, footstep vibrations are sensed by wearable accelerometers.
Real-time visual and haptic streams are delivered to a remote venue, where a large 3D LED wall and a vibration-efficient haptic floor envelop dozens of spectators.
A public trial at Expo 2025 linked sites 20 km apart: visitors watched a live dance show and conversed with performers without noticeable delay. 
\end{abstract}



\begin{teaserfigure}
  \centering
  \includegraphics[width=\textwidth]{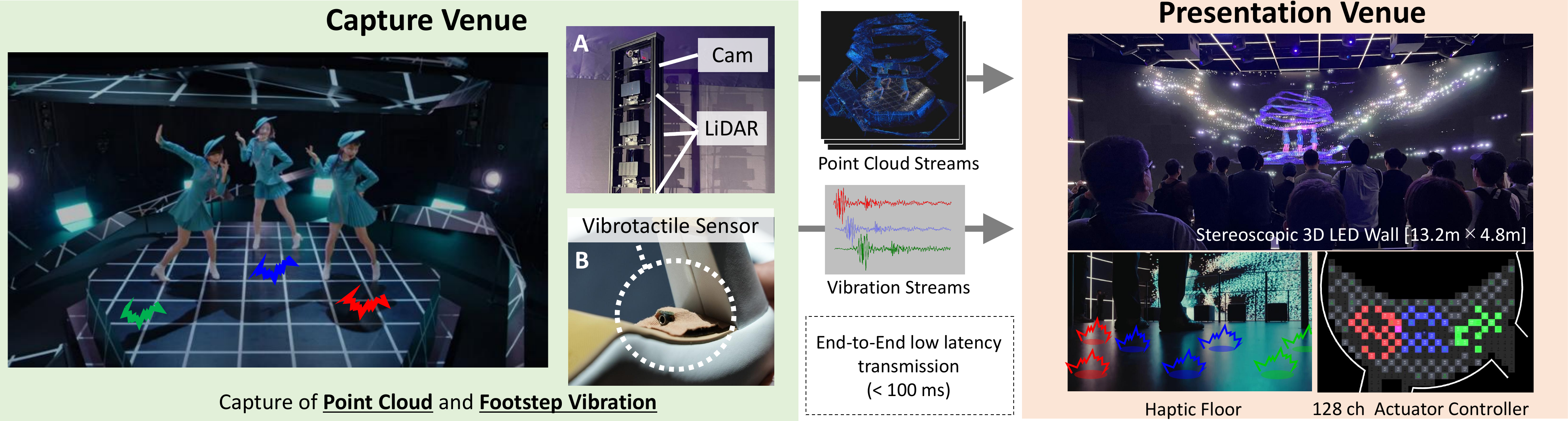}
  \caption{Performers and stage captured as dynamic point clouds; footsteps sensed as vibrations: (A) capture unit; (B) shoe‑mounted vibration sensor. Visual/haptic streams drive 3D LED wall and floor actuators, enabling shared tele‑immersion.}
  \Description{Performers and stage captured as dynamic point clouds; footsteps sensed as vibrations: (A) capture unit; (B) shoe‑mounted vibration sensor. Visual/haptic streams drive 3D LED wall and floor actuators, enabling shared tele‑immersion.}
  \label{fig1}
\end{teaserfigure}


\maketitle

\section{Introduction}
Live performances engage sight, sound, and sometimes felt vibrations; with recent low‑latency networking, such multisensory, bidirectional immersion is feasible when stage and audience are apart. 

Stage-scale, real-time 3D with haptics requires four capabilities: wide-area, multi-performer capture robust under rapid lighting changes; noise-robust footstep sensing; efficient delivery to many spectators; and low end-to-end latency.
Prior volumetric telepresence achieves real‑time capture but often relies on dense camera arrays or engineered lighting, limiting deployment on live stages \cite{OrtsEscolano16,Guo19,Zhou23}. On the haptics side, floor‑embedded microphone arrays require structural modification and numerous sensors and remain vulnerable to loud music \cite{Kusabuka24}; bespoke haptic floors increase cost and complexity \cite{Bouillot19}; and audience‑worn devices impose operational burdens at crowd scale \cite{Shen25}.

We address these needs with Dynamic Point‑Cloud Space Transmission (DPCS) and Vibrotactile‑Sound Field Presentation (VSFP). 
DPCS fuses multiple LiDARs and RGB cameras with lightweight densification and a view‑dependent depth bias to stream full‑stage scenes with <100 ms latency; 
VSFP uses shoe‑mounted accelerometers with signal conditioning to drive a propagation‑aware floor that achieves venue‑wide vibrations with far fewer actuators.

We validated the system at Expo 2025, streaming performers and stage to a remote pavilion 20 km away; 
a large 3D LED wall and 128 floor actuators reproduced the show, and a live talk confirmed natural two‑way communication with no perceptible delay.

\section{Dynamic Point-Cloud Space Transmission}
We capture wide‑area 3D scenes with multiple time‑synchronized units, each integrating three 128‑channel LiDARs ($360^{\circ}$ FoV, 10 Hz, ±3 cm) and a full‑HD global‑shutter RGB camera (Fig.\ref{fig1}(A)). 
Using the PTP, the three LiDAR sweep phases are staggered by 120°, so any azimuth is revisited every 33 ms; the camera is hardware‑triggered per sweep, yielding synchronized RGB-D at 30 fps. 

Because the LiDARs are spatially displaced, their beams hit slightly different surface points across frames; static regions therefore appear to flicker despite being motionless.

To densify static areas and suppress inter‑frame flicker at low cost, we use a lightweight four‑step, per‑frame procedure: 
(i) project each LiDAR onto the synchronized RGB to obtain a camera‑space depth image; 
(ii) detect motion by inter‑frame color differencing; 
(iii) perform a single morphological closing to form closed motion masks; 
and (iv) fuse static points from the previous two frames into the current depth while excluding moving regions.

Each RGB-D frame is transmitted as follows: RGB is encoded as JPEG, while depth is stored losslessly in 16-bit QOI.

A single GPU host decodes all streams, back‑projects them to 3D, and fuses the point clouds. 
To mitigate seams from residual inter-unit calibration errors, we apply a view-dependent depth bias $D = s(\boldsymbol{v}\cdot\boldsymbol{i})$, where $\boldsymbol{v}$ is the virtual-camera ray and $\boldsymbol{i}$ is the capture unit's optical axis, with $s = 0.05$ for our setup, tuned empirically. Subtracting $D$ before the z‑test prioritizes points whose viewing direction best matches the virtual camera, greatly reducing seams. 
This pipeline enables full‑stage streaming with <100 ms latency.

\section{Vibrotactile-Sound Field Presentation}
We implement VSFP with two elements: on‑shoe accelerometers (Fig.\ref{fig1}(B)) for robust footstep capture and a propagation‑aware vibration floor for wide‑area playback. 
Prior work \cite{Kusabuka24} captured footsteps with floor‑embedded microphones; however, that approach required extensive cabling and was vulnerable to loud music.
In contrast, our shoe-mounted accelerometers directly sense vibration at the performer's shoes, are inexpensive, work on various floor types, and transmit wirelessly, robust to noise.

Raw accelerometer signals may contain multiple peaks per step and sound muffled or boomy when directly actuated. We therefore apply: (1) a gait-periodicity gate to align bursts with individual steps, and (2) an equalization filter to shape the vibration spectrum to match the visual impression.
These steps yield natural, texture-rich haptic feedback synchronized with the visuals.

For venue‑scale delivery without special flooring, our propagation-aware floor uses two laterally offset plywood layers to enhance vibration coupling across modular raised-access panels.
This design achieves uniform coverage with about one-third the actuators of a tile-by-tile layout, reducing cost while scaling to large spaces. Measurements showed less than 6 dB attenuation at the farthest tiles, confirming effective vibration propagation.

A Unity‑based controller maps dancers’ step signal and position to actuator channels, supporting patterns like floor‑wide excitation for collective presence and localized excitation for spatial realism.

\section{System Deployment in Live Performance}
The performance linked two sites 20 km apart: a remote stage and an Expo 2025 pavilion.
At the stage, a 6.5×6.5×4.6 m platform with a 0.6 m‑high floor was surrounded by seven capture units for full‑body acquisition; wireless heel-mounted modules measured footstep vibrations. 
Visual capture remained stable under 25 color LED fixtures (12 moving lights), and four 4K stereo cameras provided live footage interleaved with the point‑cloud rendering.

The pavilion used a $165 m^{2}$ circular layout with a 13.2 × 4.8 m stereoscopic 3D LED wall (3520×1280 px/eye) arranged in a gentle 3° arc every 0.6 m, with a buffer zone for unobstructed 3D viewing. 
The 90 $m^2$ viewing area comprised 352 raised‑access panels, yet only 128 transducers were required; tiles without a dedicated transducer showed 6 dB attenuation. 
Attendance was capped at 70 spectators but the system scales via larger-area deployment of the same floor design.
Sites were connected by dedicated fiber. Each compressed stream carried 1.5 million points per frame at 30 fps, totaling 3.5 Gbps across seven units and footstep signals 24 Mbps total.

End‑to‑end DPCS latency averaged 81.3 ms, below the 150 ms one‑way threshold of ITU‑T G.114, supporting smooth interaction. 
A 4.5‑minute live show mixed streamed point clouds with 4K stereoscopic footage and CG.
The 128 floor actuators were driven by each dancer’s step signals with floor‑wide or localized patterns chosen per scene.
Spectators wore passive 3D glasses and could stand freely; a subsequent 10‑minute live talk with the remote stage confirmed smooth bidirectional communication, and the show was later replayed in the pavilion.
Social‑media analysis (Meltwater) retrieved 6,012 posts between 13 Apr–6 May: 74.5\% positive, 22.4\% neutral, 3.1\% negative. Positive comments frequently noted strong presence and feeling the footsteps, e.g., “It felt like the dancers were right here” and “I could sense every step.”

\section{Conclusion}
We developed a tele-immersive system that fuses synchronized point clouds from multiple LiDAR–camera units with lightweight densification, flicker suppression, and a view-dependent depth bias, and reproduces footstep vibrations via wearable sensors and a propagation-aware floor. In an Expo 2025 deployment, three dancers and their stage were streamed to a remote pavilion with a 90 m² floor driven by 128 actuators. The system achieved 81 ms end-to-end latency, enabling natural conversation and a strong sense of presence, demonstrating the first stage-scale integration of real-time 3D and haptic transmission.

\bibliographystyle{ACM-Reference-Format}
\bibliography{expo2025_paper}


\end{document}